\documentclass[aps,pra,twocolumn,floatfix,footinbib,notitlepage,groupaddress,showpacs,superscriptaddress]{revtex4-1}
\usepackage{graphicx,graphics,times,bm,bbm,bbold,amssymb,amsmath,amsfonts,dsfont,hyperref,xcolor,color,cancel}
\usepackage[normalem]{ulem}
\usepackage{upgreek}
\hypersetup{
    colorlinks,
    linkcolor={blue},
    citecolor={blue},
    urlcolor= {blue}
}

\newcommand{\ket}[1]{\left|{#1}\right\rangle}

\newcommand{\ketbra}[2]{|{#1}\rangle \langle{#2}|}
\newcommand{\eeqref}[1]{Eq.~(\ref{#1})}
\newcommand{\figref}[1]{Fig.~\ref{#1}}

\newcommand{\ignore}[1]{}
\makeatletter
\newcommand*{\rom}[1]{\expandafter\@slowromancap\romannumeral #1@}
\makeatother

\let\oldsqrt\sqrt
\def\sqrt{\mathpalette\DHLhksqrt}
\def\DHLhksqrt#1#2{%
\setbox0=\hbox{$#1\oldsqrt{#2\,}$}\dimen0=\ht0
\advance\dimen0-0.2\ht0
\setbox2=\hbox{\vrule height\ht0 depth -\dimen0}%
{\box0\lower0.4pt\box2}}

\DeclareFontFamily{OT1}{pzc}{}
\DeclareFontShape{OT1}{pzc}{m}{it}%
              {<-> s * [1.25] pzcmi7t}{}
\DeclareMathAlphabet{\mathpzc}{OT1}{pzc}%
                                 {m}{it}

\begin{document}

\title{Binding energy of bipartite quantum systems: Interaction, correlations, and tunneling}

\author{M. Afsary}
\affiliation{Department of Physics, Sharif University of Technology, Tehran 14538, Iran}

\author{M. Bathaee}
\affiliation{Department of Physics, Sharif University of Technology, Tehran 14538, Iran}

\author{F. Bakhshinezhad}
\affiliation{Department of Physics, Sharif University of Technology, Tehran 14538, Iran}
\affiliation{Institute for Quantum Optics and Quantum Information - IQOQI Vienna, Austrian Academy of Sciences, Boltzmanngasse 3, 1090 Vienna, Austria}

\author{A. T. Rezakhani}
\affiliation{Department of Physics, Sharif University of Technology, Tehran 14538, Iran}

\author{A. Bahrampour}
\affiliation{Department of Physics, Sharif University of Technology, Tehran 14538, Iran}

\begin{abstract}
We provide a physically motivated definition for the binding energy (or bond-dissociation) of a bipartite quantum system. We consider coherently applying an external field to cancel out the interaction between the subsystems, to break their bond and separate them as systems from which no work can be extracted coherently by any cyclic evolution. The minimum difference between the average energies of the initial and final states obtained this way is defined as the binding energy of the bipartite system. We show the final optimal state is a passive state. We also discuss how required evolution can be realized through a sequence of control pulses. We illustrate utility of our definition through two examples. In particular, we also show how quantum tunneling can assist or enhance bond-breaking process. This extends our definition to probabilistic events. 
\end{abstract}
\date{\today}

\maketitle

\section{Introduction}
\label{sec:intro}

Binding energy ($\mathcal{BE}$) or bond-dissociation energy is a prevalent concept in various branches of science such as physical chemistry, atomic physics, nuclear physics, and gravitation. Colloquially, e.g., in chemistry, $\mathcal{BE}$ is defined as the energy needed to fully decompose a composite material into its constituent elements (in a mole of material). Some examples where $\mathcal{BE}$ can be relevant are ionization of an atom, alpha particle decay \cite{takigawa2017fundamentals}, or dissociation of molecules. In addition to advanced measurement techniques, there exist numerical methods in physical chemistry---e.g., the finite-difference Poisson-Boltzmann electrostatic method---to theoretically compute $\mathcal{BE}$ for materials \cite{schapira1999prediction}.

In \textit{classical} systems, $\mathcal{BE}$ is attributed to the forces that bound elements of a composite system together \cite{rittner1951binding}. However, with the recent advent of nanotechnology and engineering small-scale systems, it seems important to revisit the concept of $\mathcal{BE}$ for systems where \textit{quantum} effects prevail \cite{schreiner2011methylhydroxycarbene, li2011quantum, richardson2016concerted,allahverdyan2004maximal}. In particular, quantum coherence and quantum correlations have a role in physical and chemical evolutions since they contribute to energy exchange and thermodynamics of quantum systems \cite{perarnau2015extractable,2016-Alipour}. Additionally, it has also been argued that quantum tunneling may be employed in some dynamical evolutions \cite{takigawa2017fundamentals} or chemical reactions in order to reduce required initial energy in some bond-breaking processes \cite{schreiner2011methylhydroxycarbene}. 

Various technical tools have been developed in order to investigate control and manipulation of quantum systems. For example, optimal control theory (OCT) introduces techniques based on variational optimization and differential geometry to obtain optimal approaches to achieve a target in quantum systems \cite{dong2010quantum,leitmann1981calculus, fleming2012deterministic,dalessandro,palao,mr-rezvani,balint2005quantum, von2012optimal,meystre2013short, absil2001vertically, harrison2016quantum,absil2001vertically, takigawa2017fundamentals}, e.g., by \textit{coherently} applying appropriate control fields such as lasers pulse trains \cite{shnitman1996experimental, blazy1980binding}. 

Here we introduce a definition for the $\mathcal{BE}$ of a quantum bipartite system and propose methods to (optimally) break a bond in a composite system. We restrict ourselves to \textit{unitary} processes during which a bond breaks due to \textit{work extraction}. We consider several scenarios for breaking a bond by external control, and discuss optimal or close-to optimal control strategies. In particular, we focus on photodissociation where a bond breaks through absorption of photons generated by suitable laser pulse trains \cite{schlemmer2015laboratory}.

This paper is organized as follows. We start by briefly reviewing, in Sec. \ref{sec:rev}, how control of a quantum system affects it. In Sec. \ref{sec:definition}, we present and motivate a definition for $\mathcal{BE}$. In Sec. \ref{sec:opt}, we obtain optimal state and coherent evolution for bond breaking. This section also includes discussions of an example of bond breaking in an atom-cavity system. We discuss the impact of quantum tunneling in bond breaking in Sec. \ref{tunnel}. The paper is summarized in Sec. \ref{sec:summary}.


\begin{figure}[bp]
\centering
\includegraphics[scale=0.27]{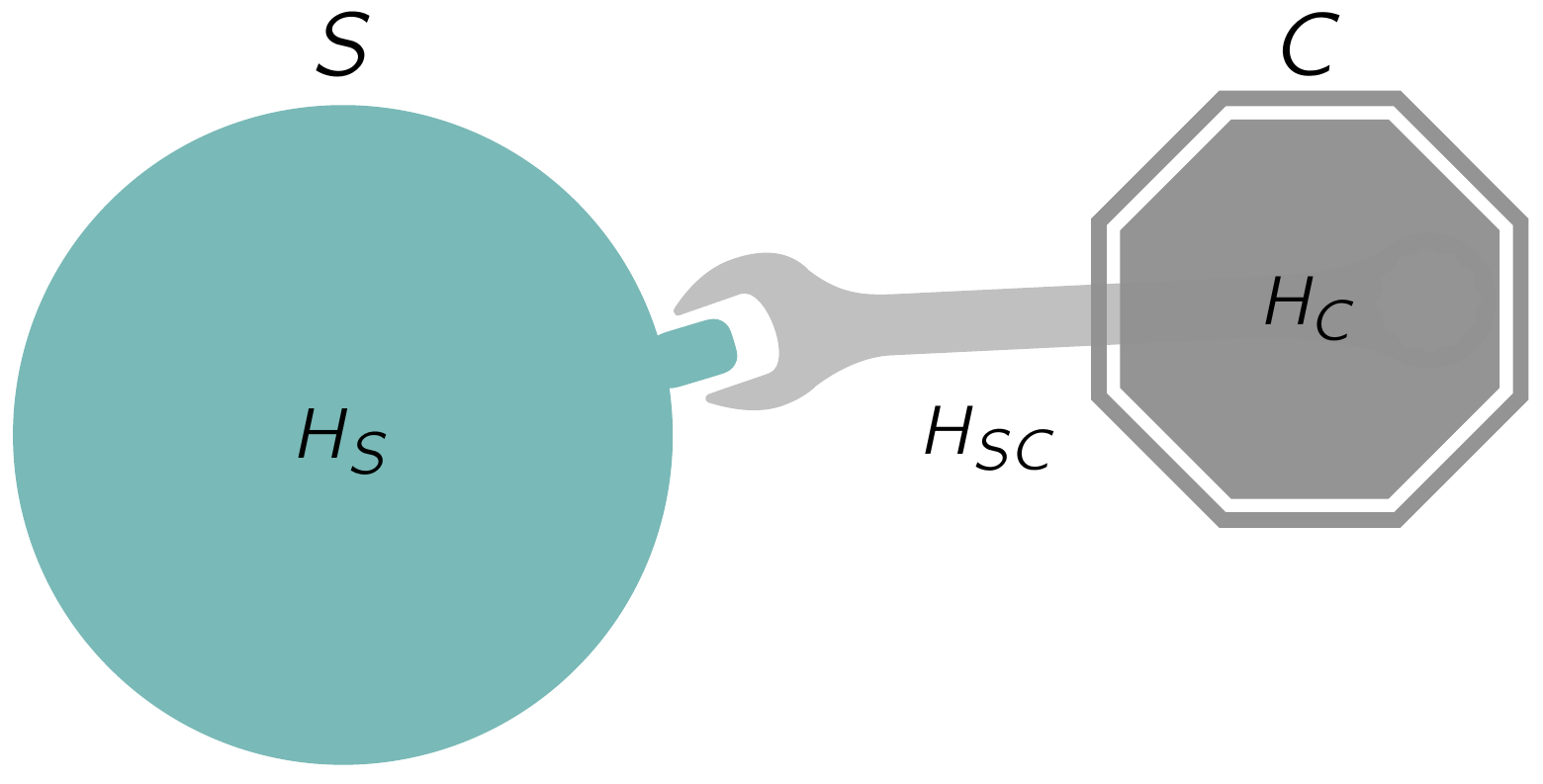}
\caption{Schematic of a system under control by another system.}
\label{fig:control}
\end{figure}

\section{Controlling a quantum system}
\label{sec:rev}

Let us assume that we manipulate a quantum system ($\mathsf{S}$) with an external control agent or apparatus ($\mathsf{C}$), which is another classical or quantum system. The Hilbert space of the composite system is $\mathpzc{H}_{\,\mathsf{SC}}=\mathpzc{H}_{\,\mathsf{S}} \otimes \mathpzc{H}_{\,\mathsf{C}}$. It is known that the total Hamiltonian of this composite system is given by
\begin{align}
H_{\mathsf{SC}}(\bm{\lambda}, \mathbf{g};t) = H_{\mathsf{S}}(\bm{\lambda};t) + H_{\mathsf{C}} + H_{\mathrm{int}}(\mathbf{g};t).
\end{align}
Here $H_{\mathsf{S}}(\bm{\lambda};t)$ indicates the system Hamiltonian, which may depend explicitly on time as well as some other structural parameters $\bm{\lambda}$ (e.g., size of the box for particle in a box). The Hamiltonian of the controller is shown by $H_{\mathsf{C}}$, which for simplicity we assume to be time-independent. The interaction Hamiltonian $H_{\mathrm{int}}(\mathbf{g};t)$ is the main player in controlling system $\mathsf{S}$, which itself may depend on time and some structural parameters $\mathbf{g}$ given by the physics of the two systems $\mathsf{S}$ and $\mathsf{C}$ and how they interact (e.g., an electron and an electric field)---see Fig. \ref{fig:control}. 

Note that although the Hilbert space after the control is $\mathpzc{H}_{\,\mathsf{SC}}$, in general interaction of the system and the controller can yield a different control-induced decomposition as $\mathpzc{H}_{\,\mathsf{SC}}=\mathpzc{H}_{\,\mathsf{S}_{1}}\otimes \mathpzc{H}_{\,\mathsf{S}_{2}}\otimes \ldots \mathpzc{H}_{\,\mathsf{C}'}$, where new physical parties may be produced and the control system may also drastically vary. Evolution of each product or party ($\ell\in\{\mathsf{S}_{1},\ldots,\mathsf{C}'\}$) is then given by a dynamical equation obtained by reducing (tracing out) the total dynamical equation \cite{book:Breuer}, 
\begin{equation}
\frac{\partial}{\partial t}\varrho_{\ell}(\bm{\lambda},\mathbf{g};t) =-i[H^{(\mathrm{eff})}_{\ell}(\bm{\lambda},\mathbf{g};t), \varrho_{\ell}(\bm{\lambda},\mathbf{g};t)] +L_{\ell}(\bm{\lambda},\mathbf{g};t),
\end{equation}
where (and henceforth throughout the paper) we have assumed $\hbar\equiv 1$. Here the density matrix $\varrho_{\ell}= \mathrm{Tr}_{\bar{\ell}}[\varrho_{\mathsf{SC}}]$, with $\bar{\ell}=\{\mathsf{SC}\}-\{\ell\}$, describes the quantum state of party $\ell$ and 
\begin{equation}
H^{(\mathrm{eff})}_{\ell} = H_{\ell} + \Delta H_{\ell},
\end{equation}
is the \textit{effective} Hamiltonian of party $\ell$, which is the Hamiltonian $H_{\ell}$ originally attributed to party $\ell$ renormalized by the correction $\Delta H_{\ell}$ due to all remaining parties ($\bar{\ell}$). This effective Hamiltonian describes the \textit{coherent} part of the dynamics. In addition to this part, there is an $L_{\ell}$ which encompasses \textit{incoherent} (i.e., nonunitary) part of the dynamics of the party which incorporates correlations and interactions with other parties \cite{book:Breuer,correlation-picture}. 

However, under some specific condition the dynamics of a controlled system can still be described \textit{coherently}. Consider the following conditions: (i) the applied control, e.g., a field, is sufficiently \textit{weak} ($\Vert H_{\mathrm{int}}(\mathbf{g})\Vert\ll \Vert H_{\mathsf{S}}(\bm{\lambda}) + H_{\mathsf{C}}\Vert$, where $\Vert\cdot\Vert$ is the standard operator norm); (ii) the control-induced decomposition of the total Hilbert space is still the same as the decomposition before control; and (iii) the change in the control system $\mathsf{C}$ is not appreciable or of interest (thus it can be simply ignored), then the contribution of the incoherent term in the dynamics of system $\mathsf{S}$ may be negligible, $\Vert L_{\ell}(\bm{\lambda},\mathbf{g})\Vert\approx 0$. Under such conditions, the action of the control on the system can be \textit{effectively} recast as a change of the system Hamiltonian as
\begin{equation}
H^{(\mathrm{eff})}_{\mathsf{S}}(\bm{\lambda},\mathbf{g};t)=H_{\mathsf{S}} (\bm{\lambda};t)+V(\mathbf{g};t),
\end{equation}
where $V$ is a Hamiltonian associated with the applied control field, acting on $\mathpzc{H}_{\,\mathsf{S}}$. In this regime, varying the system Hamiltonian, by changing $\bm{\lambda}$ in the unperturbed system Hamiltonian $H_{\mathsf{S}}$ or by changing $\mathbf{g}$ in the applied field $V$, can yield a target dynamics for the system. As a remark, note that in some sense weakness of the control also implies weakness of $V$ with respect to $H_{\mathsf{S}}$.  

It will be helpful to consider a simple physical example; interaction of light (e.g., a laser or electric field $\hat{\mathbf{E}}$) and matter (e.g., an atom) \cite{gerry2005introductory}. When the atom has only two energy levels, the field is single-mode almost at resonance with the atom ($\omega\approx \epsilon_{2}-\epsilon_{1}$), and it is sufficiently weak so that the dipole approximation suffices ($H_{\mathrm{int}}=-\hat{\bm{D}}\cdot\hat{\bm{E}}$, with $\hat{\bm{D}}$ being the dipole moment operator of the atom), the total Hamiltonian of this atom-field system can be described by the Jaynes-Cummings model,
\begin{equation}
H_{\mathsf{SC}} = H_{\mathsf{S}} + H_{\mathsf{C}}  + \textit{g}(e^{-i\omega t}\hat{a}\otimes |e\rangle \langle g| + e^{i\omega t}\hat{a}^{\dag}\otimes |g\rangle \langle e|),
\end{equation}
where $H_{\mathsf{S}}=\sum_{i} \epsilon_{i}|i\rangle\langle i|,~ i\in \{e,g\}$ is the Hamiltonian of the atom, and $H_{\mathsf{C}} = \omega (\hat{a}^{\dag}\hat{a}+1/2)$ is the field Hamiltonian, with $\hat{a}$ being the bosonic annihilation operator of the field mode.

In the coherent regime, if the field is \textit{classical}, we can say its action on the atom is given by the potential $V(\textit{g},\omega;t)=\textit{g} (e^{i\omega t}|g\rangle \langle e| + e^{-i\omega t}|e\rangle \langle g|)$, which induces transitions between the atomic energy levels and $\textit{g}$ is the coupling strength. Indeed, this approach is taken in elementary considerations of how an atom interacts with an electric field and yields the Rabi oscillation which presents the emission and absorption of the photon between atom and field periodically. Hence, it  mimics the binding energy between field source and atom \cite{gerry2005introductory,book:Sakurai}.

\section{$\mathcal{BE}$ of bipartite quantum systems}
\label{sec:definition}

Consider a composite bipartite system $\mathsf{S}$, comprised of two parts $\mathsf{A}$ and $\mathsf{B}$. The associated Hilbert spaces of the subsystems and the composite system are denoted by $\mathpzc{H}_{\,\mathsf{A}}$, $\mathpzc{H}_{\,\mathsf{B}}$, and $\mathpzc{H}_{\,\mathsf{S}}=\mathpzc{H}_{\,\mathsf{A}}\otimes \mathpzc{H}_{\,\mathsf{B}}$. The free Hamiltonians of the subsystems $\mathsf{A}$ and $\mathsf{B}$ are given by $H_{\mathsf{A}}=\sum_{i=1}^{d_{\mathsf{A}}}\epsilon^{(\mathsf{A})}_{i}|i\rangle_{\mathsf{A}}\langle i|$ and $H_{\mathsf{B}}=\sum_{i=1}^{d_{\mathsf{B}}}\epsilon^{(\mathsf{B})}_{i}|i\rangle_{\mathsf{B}}\langle i|$. The Hamiltonian of the composite system $\mathsf{AB}$ is assumed to be
\begin{equation}
H=H_{\mathrm{free}}+H_{\mathrm{int}},
\end{equation}
where $H_{\mathrm{free}}=H_{\mathsf{A}}+H_{\mathsf{B}}$ is the free Hamiltonian of the composite system, and $H_{\mathrm{int}}$ describes the interaction between the subsystems. We assume the spectral decomposition $H_{\mathrm{free}}=\sum_{\gamma=1}^{d}\epsilon_{\gamma}|\Phi_{\gamma}\rangle \langle \Phi_{\gamma}|$, where $d=d_{\mathsf{A}}d_{\mathsf{B}}=\dim(\mathpzc{H}_{\,\mathsf{S}})$, $\gamma\equiv(i,j)$ with $i\in\{1,\ldots,d_{\mathsf{A}}\}$ and $j\in\{1,\ldots,d_{\mathsf{B}}\}$, $\epsilon_{\gamma}\equiv \epsilon^{(\mathsf{A})}_{i} + \epsilon^{(\mathsf{B})}_{j}$, and $|\Phi_{\gamma}\rangle\equiv|i\rangle_{\mathsf{A}}|j\rangle_{\mathsf{B}}$ are the eigenstates of the free Hamiltonian (also known as the ``bare states"). Similarly, we assume the spectral decomposition $H=\sum_{\gamma} E^{[\mathrm{D}]}_{\gamma}|\Phi^{[\mathrm{D}]}_{\gamma}\rangle \langle \Phi^{[\mathrm{D}]}_{\gamma}|$ (where $|\Phi^{[\mathrm{D}]}_{\gamma}\rangle$ are called ``dressed states").

The instantaneous state of the composite system at any time $\varrho(t)$ can be represented as \textcolor{blue}{\cite{Mahler2010spinoscillator}}
\begin{align}
\label{den-mat}
\varrho(t)=\varrho_{\mathsf{A}}(t) \otimes \varrho_{\mathsf{B}}(t)+\chi(t),
\end{align}
where $\varrho_{\mathsf{A}}$ and $\varrho_{\mathsf{B}}$ are the reduced density matrices of the subsystems, and $\chi$ denotes correlations, classical or quantum, in the composite system. Note that $\mathrm{Tr}_{\mathsf{A}}[\chi]=\mathrm{Tr}_{\mathsf{B}}[\chi]=0$. In addition, the (``average" or ``internal") energy associated to the system is given by $U(t)=\mathrm{Tr}[H(t)\,\varrho(t)]$. 

\ignore{
First, we need to specify the parameters the binding energy may depends on. It seems that the initial state of the compound system is an important factor in the energy needed to break the bond between two subsystems: for an atom in ground state, the binding energy of the nucleus and the electrons (ionization energy) is more than the binding energy of excited atoms.
On the other hand, the interaction between two parts of the system is a considerable parameter in surveying binding energy. It is known that when the bond of two particles breaks, they cannot interact anymore. Hence, the interaction Hamiltonian should be automatically ``\textit{turned off}" in such a process. Moreover, correlations between two subsystems must be taken into account. Note that in natural molecules the binding energy is always positive and you should spend energy to detach the subsystems. We call such a coupling an \textit{endoergic bond}. On the other side, we have the \textit{exoergic bonds} for which you receive energy while breaking the bond. This kind of bond breaking works for artificial atoms such as rare-gas halide molecules in excimer lasers which are generated in the interaction of excited atoms. The decomposition of such molecules releases the energy of the excited atoms \cite{x} \textcolor{blue}{[AR: missing reference.]}
}

To dissociate parts $\mathsf{A}$ and $\mathsf{B}$, a suitable time-dependent potential $V(t)$ is applied which modifies the Hamiltonian as
\begin{align}
H\,\to\,H(t)\equiv H(V(t)),
\end{align}
such that $V(0)=0$ and $H(V(0))\equiv H$ and at the dissociation time $t_{\mathrm{f}}$ where again $V(t_{\mathrm{f}})=0$ the interaction part ($H_{\mathrm{int}}$) is turned off, i.e., $H(V(t_{\mathrm{f}}))=H_{\mathrm{free}}$. 
The energy change of the system during this process is given by
\begin{align}
U(t_{\mathrm{f}})-U(t_{\mathrm{i}})= \mathrm{Tr}[\varrho(t_{\mathrm{f}}) \, H(t_{\mathrm{f}})]- \mathrm{Tr}[\varrho(0) \, H(0)].
\label{energy-diff}
\end{align}
Here $\varrho(t)$ depends on the applied external control field $V(t)$ through the evolution equation
\begin{equation}
\frac{\partial \varrho(t)}{\partial t} =-i [H(V(t)),\varrho(t)],
\label{von-neumann}
\end{equation}
or equivalently through 
\begin{equation}
\varrho(t)=\mathpzc{U}(t)\varrho(0)\mathpzc{U}^{\dag}(t), \,\,0\leqslant t\leqslant t_{\mathrm{f}},
\label{unitary-eq}
\end{equation}
where the evolution is given by $\mathpzc{U}(t)=\mathbbmss{T}e^{-i\int_{0}^{t}H(V(s))\,\mathrm{d}s}$ and $\mathbbmss{T}$ is the time-ordering operation. This evolution for a controllable composite system of dimension $d$ belongs to the unitary group $\mathbbmss{U}(d)$.

\begin{figure}[tp]
\centering
\includegraphics[scale=0.3]{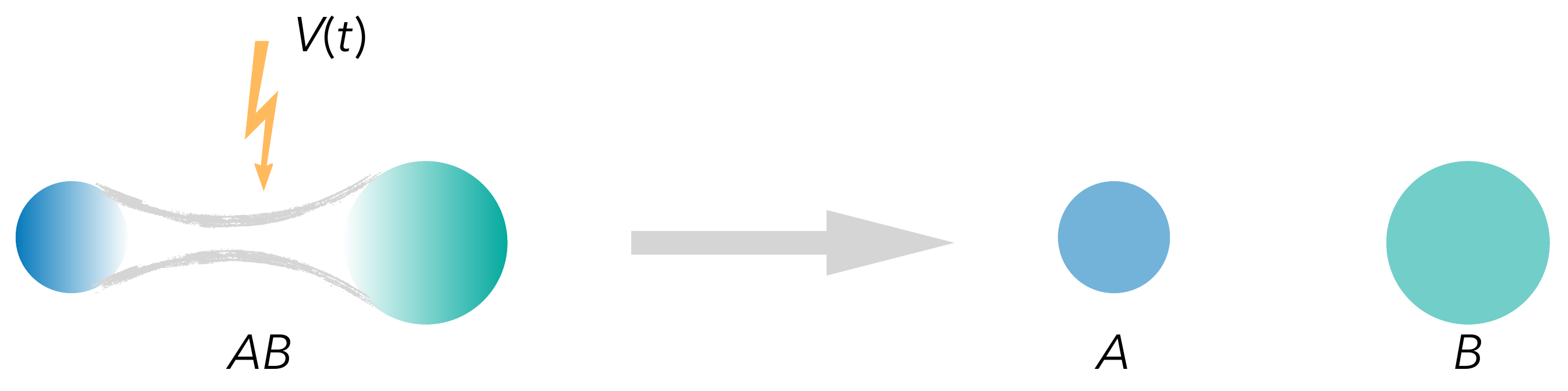}
\caption{Coherent separation of a composite system (e.g., a molecule) into its subsystems by applying an external field $V(t)$.}
\label{fig:molecule}
\end{figure}

Now, it is natural to define the $\mathcal{BE}$ as the \textit{optimal} energy required to eliminate the interaction Hamiltonian of the composite system in a \textit{coherent} manner, i.e., 
\begin{align}
\label{BE}
\Delta U_{\textsc{be}}=& \underset{t_{\mathrm{f}},V(t)}{\min} \big[\mathrm{Tr}[\varrho(t_{\mathrm{f}}) H(t_{\mathrm{f}})]-\mathrm{Tr}[\varrho(0) \, H(0)]\big]\\
=& \mathrm{Tr}[\varrho^{(\mathrm{opt})}(t^{(\mathrm{opt})}_{\mathrm{f}}) H_{\mathrm{free}}] - \mathrm{Tr}[\varrho(0) (H_{\mathrm{free}}+ H_{\mathrm{int}})].\nonumber
\end{align}
To lighten the notation, henceforth we denote the optimal time $t_{\mathrm{f}}^{(\mathrm{opt})}$ with $t_{\mathrm{f}}$ and the optimal state $\varrho^{(\mathrm{opt})}(t_{\mathrm{f}})$ with $\varrho(t_{\mathrm{f}})$. 

Several remarks are in order. (i) The minimization over time is important because if bond breaking takes too long, the composite system may experience decoherence due to environmental interactions. This optimization can be performed by employing \textit{Pontryagin's maximum principle} \cite{leitmann1981calculus}, which enables us to find the optimal control field for minimum time. (ii) The last term in Eq. (\ref{BE}) is the initial internal energy of the composite system, which is fixed and given; hence, we only need to vary the final state and the Hamiltonian to find the $\mathcal{BE}$---the energy needed for dissociation. Note that the sole result of this evolution should be effectively neutralizing the interaction Hamiltonian. That is, the final Hamiltonian should be equal to the free Hamiltonian of the system, $H(t_{\mathrm{f}})=H_{\mathrm{free}}$. This yields that the average energy of the final state is 
\begin{align}
U(t_{\mathrm{f}}) =\mathrm{Tr}[\varrho_{\mathsf{A}}(t_{\mathrm{f}}) H_{\mathsf{A}}]+\mathrm{Tr}[  \varrho_{\mathsf{B}}(t_{\mathrm{f}}) H_{\mathsf{B}}].
\end{align}
(iii) It is evident that the value of $U(t_{\mathrm{f}})$ is independent of the correlations $\chi$. Thus with this definition of the $\mathcal{BE}$, non-interacting subsystems may still be correlated. Nevertheless, the optimization of Eq. (\ref{energy-diff}) guarantees that the final state of the system is a \textit{passive state}, from which by definition it is impossible to extract any \textit{work} in a \textit{cyclic process} \cite{allahverdyan2004maximal}. That is, all work-generating correlations have already been eliminated from the final state, and thus the residual correlations can only yield \textit{heat}. To remove such leftover correlations one should employ strategies which may require sophisticated handling of the state in a nonunitary fashion.

\section{Optimizations}
\label{sec:opt}

\subsection{Optimal final state $\varrho(t_{\mathrm{f}})$}
\label{sec:theory}

As remark in the previous section, the optimization (\ref{BE}) can be performed by varying $\varrho(t_{\mathrm{f}})$ over the achievable orbit of $\varrho(0)$ via unitary transformations. The kinematical extremum of $U(t_{\mathrm{f}})$ is determined by the eigenvalues of $H_{\mathrm{free}}$ as well as the eigenvalues of $\varrho(t_{\mathrm{f}})$.

We recall that the evolution of the state $\varrho(t)$ is unitary, given by Eq. (\ref{unitary-eq}), where $\mathpzc{U}(t)\in\mathbbmss{U}(d)$. To have an extremum for the final energy $U(t_{\mathrm{f}})=\mathrm{Tr} [\varrho(t_{\mathrm{f}}) H_{\mathrm{free}}]$, it is necessary that the final density matrix $\varrho(t_{\mathrm{f}})$ commute with $H_{\mathrm{free}}$; that is, $\varrho(t_{\mathrm{f}})$ should be diagonal in the eigenbasis of $H_{\mathrm{free}}$ \cite{allahverdyan2004maximal},
\begin{align}
\varrho(t_{\mathrm{f}})=\textstyle{\sum_{\gamma}} p_{\gamma} \ketbra {\Phi_{\gamma}}{\Phi_{\gamma}}.
\label{rrho}
\end{align}
Here the probabilities $p_{\gamma}$s are the eigenvalues of the initial density matrix $\varrho(0)$. The maximum and minimum values of $U(t_{\mathrm{f}})$ belong to the finite set $\mathpzc{S}=\{\mathbf{p}\cdot \bm{\epsilon}:\, \mathbf{p} \in \Pi(\mathbf{p}) ,\,\bm{\epsilon} \in \Pi(\bm{\epsilon}) \}$, where $\Pi(\mathbf{x})$ denotes the set of all permutations of $\mathbf{x}=(x_1, x_2, \ldots , x_d)\in\mathbbmss{R}^{d}$. The maximum of the set $\mathpzc{S}$ corresponds to the case where both vectors $\mathbf{p}$ and $\bm{\epsilon}$ are nondecreasing or nonincreasing, and its minimum is obtained when either of them are nondecreasing ($x^{\uparrow}_{\gamma} \leqslant x^{\uparrow}_{\gamma+1},\,\forall\gamma$) while the other one is nonincreasing ($x^{\downarrow}_{\gamma} \geqslant x^{\downarrow}_{\gamma+1},\,\forall\gamma$), 
\begin{equation}
\mathbf{p}^{\downarrow}\cdot\bm{\epsilon}^{\uparrow}= \mathbf{p}^{\uparrow}\cdot\bm{\epsilon}^{\downarrow}\leqslant\mathbf{p}\cdot\bm{\epsilon} \leqslant \mathbf{p}^{\downarrow}\cdot\bm{\epsilon}^{\downarrow}.
\end{equation}
Thus, minimizing the final energy $U(t_{\mathrm{f}})$ leads to the \textit{passive state} which is in the form of Eq. (\ref{rrho}), where $p_{\gamma}\geqslant p_{\gamma+1}$ with $\gamma$s ordered such that $\epsilon_{\gamma}\leqslant \epsilon_{\gamma+1}$, $\forall \gamma$. As a result, we obtain
\begin{align}
\label{Fin-ener}
\Delta U_{\textsc{be}} & = \textstyle{\sum_{\gamma}}p_{\gamma} \epsilon_{\gamma} - U(0).
\end{align}
\subsection{Optimal evolution $\mathpzc{U}(t)$}
\label{sec:cal}

Here we discuss general unitary evolutions of arbitrary initial states towards desired final states by employ OCT techniques. 
We start with simple cases:

(i) \textit{Maximally-mixed initial state}: Consider the initial state $\varrho(0)=\mathbbmss{I}/d$. Because of the unitarity of the evolution, this state remains unchanged during the evolution. 

(ii) \textit{Pure initial state}: This initial state results in a pure passive state which is the ground state of the final dissociated system. If we denote the initial state of the composite system with $\ket{\Psi_{\alpha}}$, the corresponding unitary transformation to the ground state of the dissociated system ($\ket{\Phi_1}$) becomes
\begin{align}
\mathpzc{U}^{(\alpha)}(t_{\mathrm{f}})=\ketbra{\Phi_{1}}{\Psi_{\alpha}}+ \textstyle{\sum_{\gamma=2}^{d }} \ketbra{\Phi_{\gamma}}{\Psi_{\alpha(\gamma)}}\label{e16},
\end{align}
where $\ket{\Psi_{\alpha(\gamma)}}$s are states orthonormal to $\ket{\Psi_{\alpha}}$, and $\alpha(\gamma)$ has a one-to-one and regular relation with $\gamma$. Since \eeqref{e16} is independent of the transformation path it is not uniquely identified. The orthogonal vectors to $\ket{\Psi_{\alpha(\gamma)}}$ are in a $(d-1)$-dimensional subspace of the $d$-dimensional space, thus infinite sets of orthogonal sub-basis $\{\ket{\Psi_{\alpha(\gamma)}}: \gamma=2,\ldots,d-1\}$ can be found. The optimization process includes calculation of the unitary transformation $\mathpzc{U}(t)$ with minimum dissociation time $t_{\mathrm{f}}$.

(iii) \textit{Thermal initial state}: In the dressed-state basis, this initial state is represented by
\begin{align}
\varrho(0)=(1/Z^{[\mathrm{D}]}) \textstyle{\sum_{\alpha}} e^{-\beta E_{\alpha}^{[\mathrm{D}]}}\ketbra{\Phi^{[\mathrm{D}]}_{\alpha}}{\Phi^{[\mathrm{D}]}_{\alpha}}
\label{e17},
\end{align}
where $\beta=1/(k_{\mathsf{B}}T)$ is the inverse temperature (with $k_{\mathsf{B}}$ as the Boltzmann constant), and $Z^{[\mathrm{D}]}=\sum_{\alpha} e^{-\beta E_{\alpha}^{[\mathrm{D}]}}$ is the partition function of the composite system in the dressed basis. The optimal final state becomes 
\begin{align}
\varrho(t_{\mathrm{f}})=(1/Z^{[\mathrm{D}]})\textstyle{\sum_{\gamma}} e^{-\beta E_{\gamma}^{[\mathrm{D}]}} \ketbra{\Phi_{\gamma}}{\Phi_{\gamma}},
\label{e18}
\end{align}
and the optimal unitary transformation is 
\begin{equation}
\mathpzc{U}^{(\alpha)}(t_{\mathrm{f}})=\textstyle{\sum_{\gamma} }\ketbra{\Phi_{\gamma}}{\Phi^{[\mathrm{D}]}_{\alpha(\gamma)}}.
\end{equation}
It is evident that this state differs from the thermal state in the bare basis.

Since the $\mathcal{BE}$ and the corresponding unitary transformation are obtained, one only needs to obtain the optimal control potential in minimum time. A remark is in order here. After removing the interaction Hamiltonian $H_{\mathrm{int}}$, we shall argue OCT techniques that the optimal unitary transformation for reaching a desired passive state can be determined by a proper laser pulse train---see appendix \ref{App. Min time}. However, as we later argue in Sec. \ref{tunnel}, in some particular dissociation processes related to a quantum tunneling and/or photo-ionization, the laser pulse controlling can also lead to removing the interaction Hamiltonian. In fact, in the tunneling case, the dissociation is enhanced by the quantum tunneling effect.

The unitary transformations $\mathpzc{U}(t_{\mathrm{f}})$ are from unitary group $e^{\mathpzc{L}}$ generated by Lie algebra $\mathpzc{L}$ defined by the Hamiltonian of the system. The dynamics of the evolution $\mathpzc{U}(t_{\mathrm{f}})$ obeys the Schr\"{o}dinger equation and depends on a control potential $V(t)$. The transformation $\mathpzc{U}(t_{\mathrm{f}} )$ is reachable kinematically for some control potential. Here, we focus on attainable controls that can be realized by a suitable laser pulse train. In this method, the upper bound on the applied field $V(t)$ is often limited by the laser power, and its lower bound is determined by the intensity modulator extinction ratio ($M^{-}_{ij}/M_{ij}^{+}$ as defined below). The rise and fall time of the potential switching is also limited by the frequency response of the laser intensity modulator ($N^{\mp}_{ij}$), which is the frequency which determines how fast one can change the laser intensity. Thus, for a pulse train which creates different dipole interactions between levels $|\Phi^{[\mathrm{D}]}_{i}\rangle$ and $|\Phi^{[\mathrm{D}]}_{j}\rangle$ ($V_{ij}$), we have the following constraints:
\begin{align}
M^{-}_{ij}\leqslant  V_{ij}(t)\leqslant  M^{+}_{ij}, \label{limits-1}\\
N^{-}_{ij} \leqslant  \frac{\mathrm{d}V_{ij}(t)}{\mathrm{d}t} \leqslant  N^{+}_{ij}. \label{limits-2}
\end{align}

The cost function in this optimal control problem is the time minimization 
\begin{equation}
t_{\mathrm{f}}= \textstyle{\int_{0}^{t_{\mathrm{f}}}} \mathrm{d}t,
\end{equation}
subject to the dynamical equation (\ref{von-neumann}) and with some other constraints. Time minimization of this optimal process can be obtained by \textit{Pontryagin's maximum principle}. This principle states that at any instant of time, the optimal control must maximize the corresponding system ``control Hamiltonian" $\mathbbmss{H}$. This Hamiltonian is given by introducing conjugate variables $\lambda_n,\lambda_{n\neq 0}^{\prime};~n\in\{0,(i,j)\}$ and $i,j\in\{1,\ldots,d\}$ in the following form:
\begin{eqnarray}
\mathbbmss{H}=\lambda_0 f_0+ \textstyle{\sum_{i,j=1}^{d,d}} (\lambda_{ij} f_{ij} + \lambda'_{ij} R_{ij}),
\label{HC}
\end{eqnarray}
where $f_{nm}$s are the elements of the left-hand side of Eq. (\ref{von-neumann}) in the dressed states, and $R_{nm}$ as control parameter of control Hamiltonain is the $nm$th element of the time derivative of the control potential $V$. At first glance, according to Eq. (\ref{von-neumann}), the elements of $V(t)$ seem to be control parameters of the system. However, since practically jump with infinite tilt is impossible, rather than $V_{ij}$, the modulation bandwidth of the laser pulses $R_{ij}$ are the more suitable control parameters. Since the control Hamiltonian is linear versus the control parameters $(\mathrm{d}/\mathrm{d}t)V_{ij}(t) = R_{ij}$s, according to Pontryagin's maximum principle, the control $R_{ij}$s are of the \textit{bang-bang} type \cite{fleming2012deterministic}. More rigorously, one can see that $\mathbbmss{H}=A+\sum\sigma_{ij} R_{ij}$ is maximized when $R_{ij}$ acquires its maximum or minimum based on the sign of the $\sigma_{ij}$. The explicit form of $A$ and $\sigma_{ij}$ can be easily derived by Eq. (\ref{HC}). Hence, the optimal control problem is reduced to a \textit{two-point (initial and final) boundary value problem}. This considerable reduction makes the control problem amenable to laser pulses to steer the system from its initial state to the desired target state in \textit{minimum time} \cite{schirmer2002constructive}.

\subsection{Example: Atom-cavity}

We now consider an examples where breaking the bond \textit{releases} energy and the initial state of the system is pure, thus the correlation removing is plausible. 

Consider a system consisting of a two-level atom and a cavity interacting with the Jaynes-Cummings Hamiltonian,
\begin{equation}
H=\frac{1}{2}\omega_{\mathsf{A}} \, \sigma_{z}+\frac{1}{2} \omega_{\mathsf{B}} \, \hat{a}^{\dagger}\hat{a}+ \textit{g} \, (\sigma_{+} \otimes \hat{a}+\sigma_{-}\otimes  \hat{a}^{\dagger})\label{e20},
\end{equation}
where $\sigma_{z}$ is the $z$-Pauli matrix, $\sigma_{\pm}=\sigma_{x}\pm i \sigma_{y}$ (with $\sigma_{x}$ and $\sigma_{y}$ being the other Pauli matrices), $\hat{a}$ ($\hat{a}^{\dagger})$ is the annihilation (creation) operator of the cavity, $\omega_{\mathsf{A}}$ is the energy gap of the atom, $\omega_{\mathsf{B}}$ is the resonance frequency of the cavity, and $\textit{g}$ is the coupling strength. Note that unexcited atom-cavity system experiences no interaction, hence no binding energy---a case which may appear in rare gas halogenide molecules \cite{rhodes1984excimer}. Here we assume the atom-cavity molecule in the strong coupling regime and that only one photon contributes to the evolution \cite{liu2018cavity}. Thus, the eigenstates of this Hamiltonian (dressed states) are limited to $\{\ket{0,g}, \ket{\pm}, \ket{1,e}\}$, where 
\begin{align}
\ket{+}&=\cos{\phi}\ket{0,e}+\sin{\phi}\ket{1,g}, \label{dress-1}\\
\ket{-}&=-\sin{\phi}\ket{0,e}+\cos{\phi}\ket{1,g},
\label{dress-2}
\end{align}
with $\tan(\phi/2)=2\textit{g}\,(\omega_{\mathsf{A}}-\omega_{\mathsf{B}})$ \cite{gerry2005introductory}. The atom-cavity system prepared in either of the dressed states remains there forever unless the interaction is interrupted. Considering that the initial system state to be the non-passive state $\ket{+}$ or $\ket{-}$, the atom-cavity molecule dissociation occurs when both the interaction and the quantum correlation (here entanglement) are switched off. Assuming the atom is trapped in the cavity by an optical tweezer \cite{stuart2014manipulating}.
\begin{figure}[tp]
\includegraphics[scale=.3]{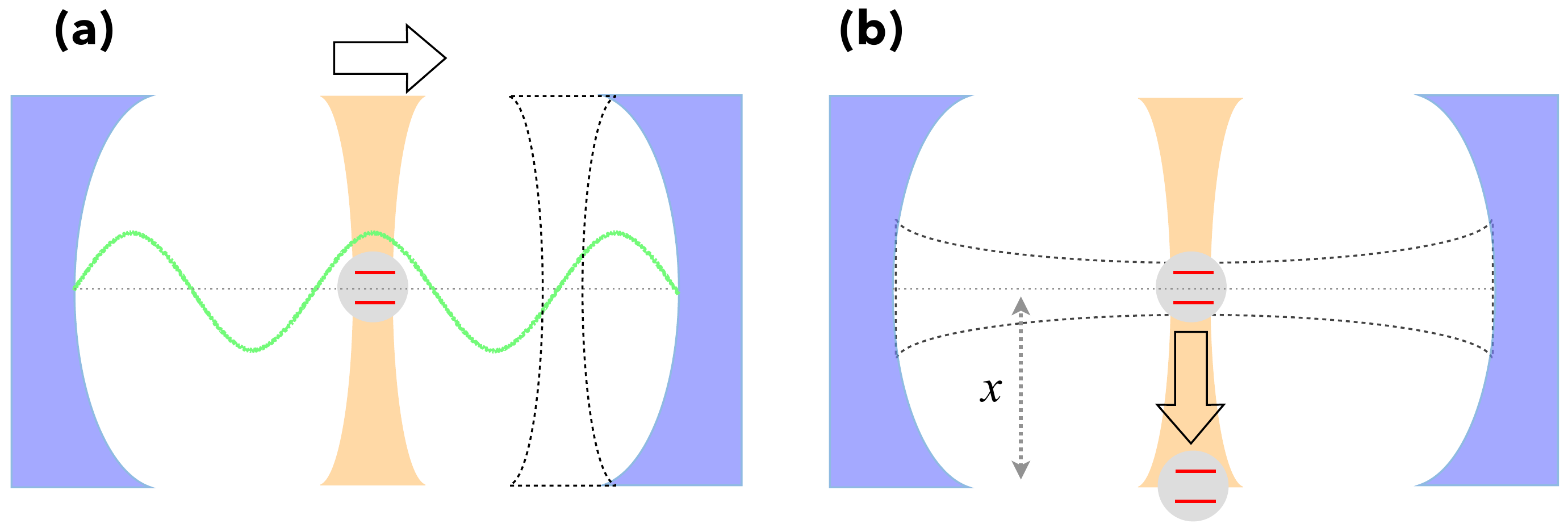}
\caption{(a) By changing the position of the optical tweezer the atom is moved to a node in the cavity, where the interaction Hamiltonian is off. (b) By turning off the optical tweezer the atom leaks out of the cavity.}
\label{cavity}
\end{figure}
By properly changing the optical tweezer's beam waist position with respect to the trapped atom's position, the atom can gain a desired velocity after switching off the optical tweezer and thus will exit the cavity \cite{nussmann2005submicron}. As depicted in \figref{cavity}, along the cross section of the cavity center the coupling strength is almost constant. If one adjusts the velocity of the atom such that $\phi(\tau)=n\pi$ for $n\in\mathbbmss{N}$ ($\tau=x/v$ is the flying time through the cavity and $v$ is the velocity) for the initial states $\ket{+}$ ($\ket{-}$), the final state of the atom-cavity becomes the bare state $\ket{0,e}$ ($\ket{1,g}$), respectively [see Eqs. (\ref{dress-1}) and (\ref{dress-2})]. This state still needs to be passivated. In the case of $\ket{1,g}$, by employing a proper pulse on the atom, the passive state $\ket{0,g}$ can be generated \cite{PhysRevLett.57.1688}; in the case of $\ket{0,e}$, the photon can escape from the cavity by changing the cavity resonance frequency, e.g., by activation of a saturable absorber in the cavity. As a result, this scenario can lead to the final passive state, which is the requirement of the bond breaking of the atom-cavity system. 

\section{Tunneling-induced bond breaking}
\label{tunnel}

\subsection{General considerations}



In addition to active control by laser pulses, it has also been demonstrated that ``quantum tunneling" may be an effective phenomena in controlling chemical reactions and molecular dissociation \cite{schreiner2011methylhydroxycarbene}. For example, photodissociation of the formaldehyde $\mathrm{H_{2}CO}$ molecule by employing quantum tunneling effect has already been reported in Ref. \cite{gray1981tunneling}. This molecule absorbs a UV-Vis photon to get excited to its upper electronic level (called ``$S_1$"), then it experiences a non-radiative emission to the upper vibrational levels of the lower electronic state (called ``$S_0$"). Now, the electron has the chance to tunnel through the potential barrier and thus the molecule is decomposed to $\mathrm{H}_{2}+ \mathrm{CO}$. Another case in which quantum tunneling results in bond breaking is the $\alpha$-decay event---see Fig. \ref{well} and Ref. \cite{takigawa2017fundamentals}. 

In some molecules attractive and repulsive forces may result in a potential barrier and tunneling effect. Alternatively, one may employ an external field, such as electrostatic and optical radiation fields, to induce a potential barrier in a bipartite system to control decomposition rate of the system. For example, electron emission may be induced by tunneling from a conductor surface in a high electric field \cite{raizer1991gas}. In this case, the required work for the potential reconfiguration should also be taken into account in the calculation of the $\mathcal{BE}$.  

Figure \ref{well} shows a typical potential barrier. Energy levels of systems with finite-width barrier can be divided to three groups: (i) bound levels, for which the tunneling rate is zero, (ii) tunneling levels with finite tunneling rate, and (iii) unbounded levels, where the tunneling probability is one. Tunneling transitions in Fig. \ref{well} are of the tunneling/decay group, where the decay transition is responsible for the depopulation of higher levels radiatively or nonradiatively. Depending on the ratio of the tunneling and decay rates of the level, the following behaviors can be discerned:

\begin{figure}[tp]
\centering
\includegraphics[scale=0.29]{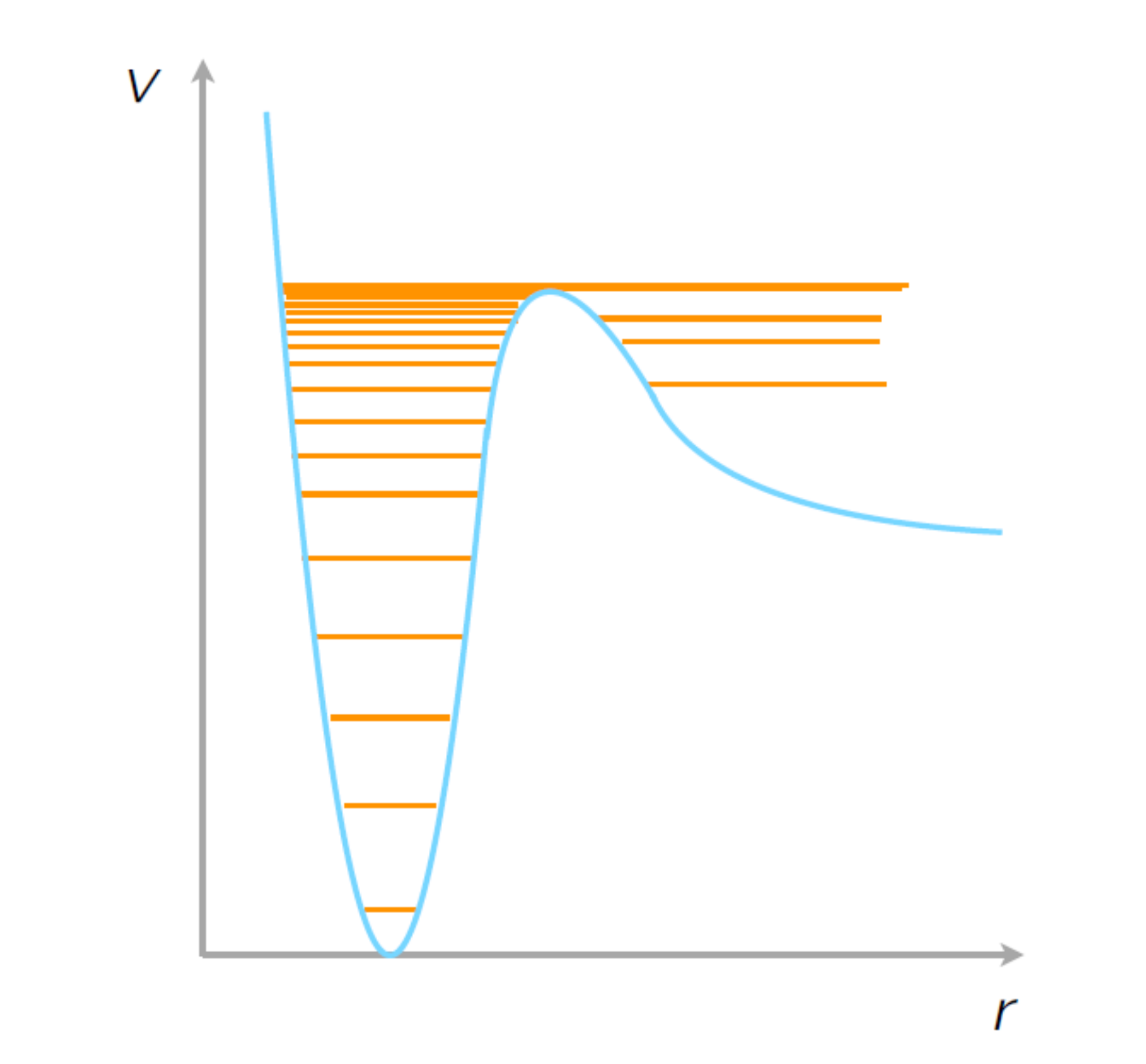}
\caption{A typical potential well with tunneling effect. A hallmark example of the latter is the ``$\alpha$-decay" event in particle physics \cite{takigawa2017fundamentals}.}
\label{well}
\end{figure}

(i) In a bipartite system with long tunneling time relative to the decay rate, transition to bound states is faster than the tunneling-induced decomposition process. In this case, if tunneling does not occur, the multi-step excitation to tunneling states should be performed as long as the decomposition can happen. For simplicity, here we only consider radiative transitions, which implies that there is no energy dissipation. Thus, populating unbounded levels may energetically cost more than multi-step excitation of the tunneling levels. Note that this condition is not dominant, because by modification of the width of the potential barrier the tunneling time can be arbitrarily reduced. Moreover, putting the molecule in a suitable cavity could increase the decay time.

(ii) When the tunneling rate is greater than the decay rate, dissociation of the molecule will be observed before the transition to the bound states.

Note that in both cases, after the tunneling process the linear momentum of the excited state is precisely determined. Due to the uncertainty relation, the position can have large uncertainty; hence, the interaction will practically vanish in the molecular dissociation process.

Since in this paper we have assumed the system to be subject to unitary evolutions, the initial and final states should have the same number of populated energy levels. Thus, if the number of the tunneling levels is less than the number of the populated levels in the initial state, one may need a multi-step excitation process until tunneling can happen. 

\begin{figure}[tp]
\centering
\includegraphics[width=0.7\linewidth]{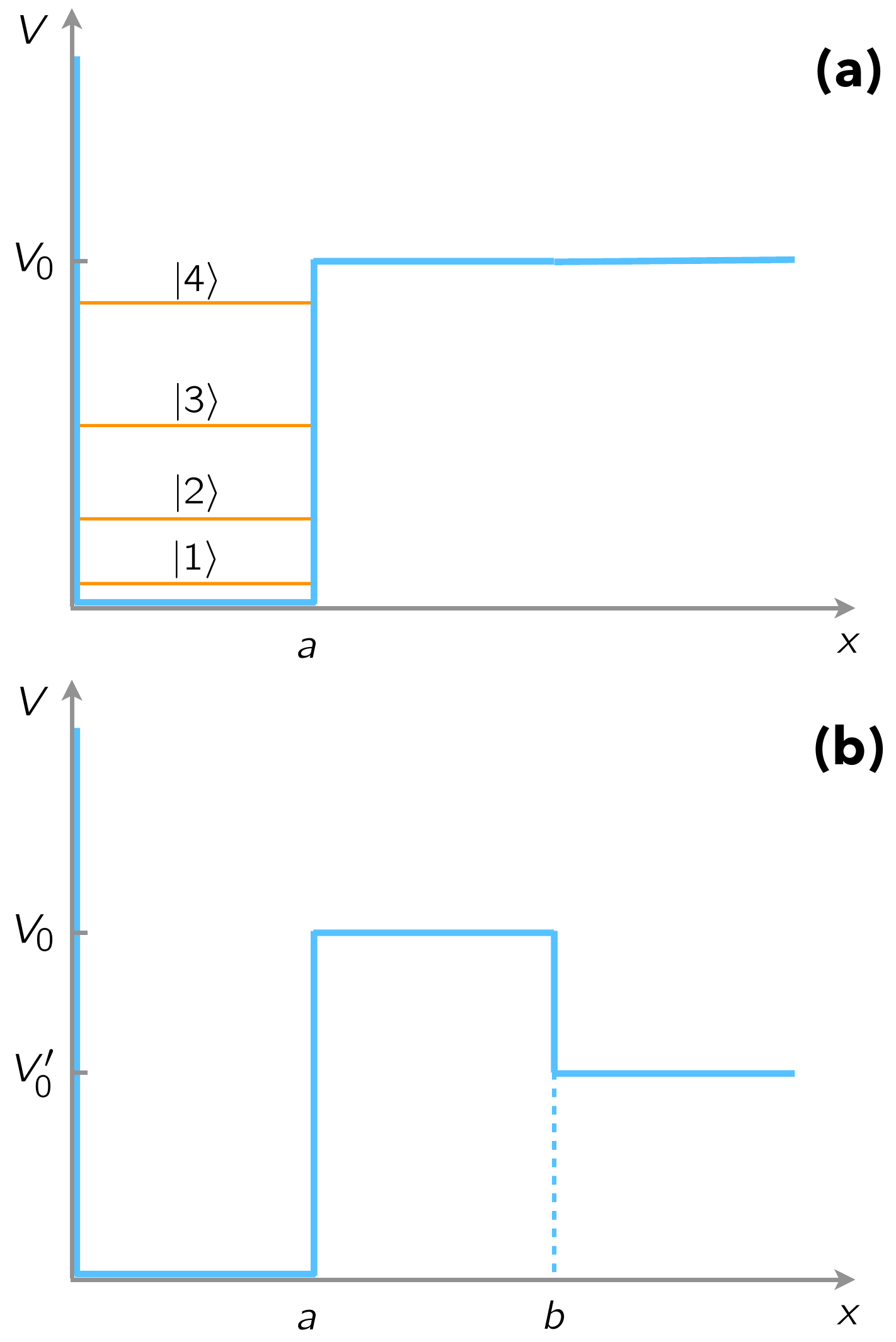}
\caption{(a) A potential step $V(x)$ with well width $a$ and height $V_{0}$, for $x\geqslant a$. There are $4$ bound states ($|k\rangle$, $k=1,2,3,4$), and no tunneling occurs here. (b) Modified potential barrier where $V(x)=V_{0}$ for $a\leqslant x\leqslant b$ and $V(x)=V'_{0}<V_{0}$ for $x>b$. Here, $a=2.62 \times 10^{-10}\,\mathrm{m}$, $b=2.80 \times 10^{-10}\,\mathrm{m}$, $V_{0}=80\,\mathrm{eV}$, and $V'_{0}=42\, \mathrm{eV}$, 
$E_{1}=4.2\,\mathrm{eV}$, $E_{2}=18.9\,\mathrm{eV}$, $E_{3}=42\,\mathrm{eV}$, and $E_{4}=72.3\,\mathrm{eV}$.}
\label{square well}
\end{figure}

\subsection{Example}
\label{subsec:tun}

Consider an electron of mass $m_{e}$ in the step potential depicted in Fig. \ref{square well} (a). The energy levels $E$ of this potential can be obtained readily by solving the equation $\tan(\sqrt{2m_{e}E} a)=-\sqrt{E/(V_{0}-E)}$ 
\cite{harrison2016quantum}. For specificity, we choose $V_0=80\, \mathrm{eV}$ and $a= 2.62 \times 10^{-10}\, \mathrm{m}$, which gives only $4$ bound states. We also take the initial state of the system in the following mixed state with no coherence:
\begin{align}
\varrho_{\mathrm{i}}=\alpha_1 \ketbra{1}{1}+ \alpha_2 \ketbra{2}{2},
\end{align}
where $\alpha_1 > \alpha_2\geqslant 0$ and $\alpha_1+\alpha_2=1$. For a quantum potential well in this shape quantum tunneling is not allowed in any energy level. Thus we apply an external control such that it only changes the potential for $x>b$ by producing a finite-width barrier---to keep simplicity, we approximate this modification as in Fig. \ref{square well} (b). The barrier width is designed such that there exist two tunneling states in the system; we choose $b=2.8 \times 10^{-10}\, \mathrm{m}$ and $V'_{0}=42\, \mathrm{eV}$.

Note that in some cases, the system should be excited to upper levels; $\varrho_{\mathrm{i}} \to \varrho'_{\mathrm{i}}$. Since the evolution is unitary, the excited state $\varrho'_{\mathrm{i}}$ has the same dimension as that of the initial one. Let us denote the number of no-tunneling and tunneling levels, respectively, with $n_{\mathrm{nt}}$ and $n_{\mathrm{t}}$. When $n_{\mathrm{t}}\geqslant  n_{\mathrm{nt}}$, the excited state $\varrho'_{\mathrm{i}}$ is diagonal with the same diagonal elements as in $\varrho_\mathrm{i}$. When $n_{\mathrm{t}}<n_{\mathrm{nt}}$, the $\varrho'_{\mathrm{i}}$ can be written versus the upper $n_{\mathrm{t}}-n_{\mathrm{nt}}$ no-tunneling levels and $n_{\mathrm{nt}}$ tunneling levels. In such states the decomposition is a multi-step procedure. Overall, this is the initial state $\varrho_{\mathrm{i}}$ that determine which scenario applies. 

In our case, $n_{\mathrm{t}}=n_{\mathrm{nt}}$ and $\varrho'_{\mathrm{i}}$ is written versus all tunneling state basis. To find the best configuration of the excited state $\varrho'_{\mathrm{i}}$, the tunneling probability of each tunneling level is needed. This probability, given by $P=e^{-2\int_a^b\sqrt{2m_{e}(V_0-E)}\mathrm{d}x}$ in WKB approximation \cite{harrison2016quantum}, for the first tunneling state [$|3\rangle$ in Fig. \ref{square well} (a)] and the second tunneling state [$|4\rangle$ in Fig. \ref{square well} (b)] can be obtained as $P_3=0.15$ and $P_4=0.40$, respectively. The tunneling time of these two levels can also be calculated. Using WKB \cite{harrison2016quantum, tanizawa1996quantum, kelkar2017electron}, we obtain $\tau_3=0.65\times 10^{-17}\,\mathrm{s}$ for the level $|3\rangle$ and $\tau_4= 1.31 \times 10^{-17}\,\mathrm{s}$ for the level $|4\rangle$. In these calculations, the tunneling rate is defined as the inverse of the product $P(2A/v)$ with $v$ the speed of the tunneling particle. As it is clear the tunneling time for both levels is relatively smaller than the decay time of the system (which is, e.g., of the order of nano second for Hydrogen).

Although the tunneling probability from the upper tunneling level ($|4\rangle$) is higher, the tunneling time of the lower tunneling state ($|3\rangle$) is sufficiently short that we do not need to force the system to the upper tunneling level. With these considerations, the system should be excited to
\begin{align}
\varrho'_{\mathrm{i}}=\alpha_1 \ketbra{3}{3}+ \alpha_2 \ketbra {4}{4}.
\end{align}
It is evident that with this choice less energy is needed to decompose the system. The corresponding unitary evolution is 
\begin{align}
\mathpzc{U}=\ketbra{3}{1}+\ketbra{4}{2}+\ketbra{1}{3}+\ketbra{2}{4}.
\label{ev}
\end{align}

Now, we want to drive the system in the optimal path which satisfies this unitary evolution and reaches the desired final state. To do so, we employ laser pulses based on OCT methods \cite{schirmer2001limits,schlemmer2015laboratory}. 
Specifically, we employ the group decomposition method of Ref. \cite{schirmer2002constructive}---see also appendix \ref{App. Min time} for a brief review---and Pontryagin's method to determine the optimal control path in \textit{minimum} time $t_{\mathrm{f}}$. In the group decomposition, the unitary operator is decomposed into a product of operators each of which is illustrative of a laser pulse \cite{ramakrishna2000explicit}. Then, the phase and the total energy of a sequence of laser pulses that drive the system through the optimal path to the desired state is calculated through Eq. (\ref{decompose})---and appendix \ref{App. Min time}. Through the method of Ref. \cite{schirmer2002constructive}, one can construct the optimal pulse sequence by finding appropriate pulses each of which causes a unitary transition to an upper level. 

Our numerical calculations shows that the pulse train for the evolution of $\varrho_i$ to $\varrho'_i$ should be applied in the following sequence
\begin{align}
\mathpzc{U}=\mathpzc{U}_{23}(t_4)\,\mathpzc{U}_{34}(t_3)\,\mathpzc{U}_{12}(t_2)\, \mathpzc{U}_{23}(t_1),
\label{e23}
\end{align}
where $\mathpzc{U}_{ij}(t_k)$ induces the dipole transition between the $i$th and $j$th levels of the (composite) system in the time interval $[t_{k-1},t_k]$, which represents the duration of the pulse. Details of the calculations of the pulse durations can be found in appendix \ref{App.example calculations}.
 
The laser pulses shapes are generated by intensity modulators where the slope of intensity increasing or decreasing are limited by the intensity modulator bandwidth. For numerical calculations, we consider a modulator with the rate $0.1\,\mathrm{GHz}$ (This is a lower bound for the modulation bandwidth, the calculations may be done with bigger ones). Furthermore, the laser intensities are limited by the laser sources, which for this example we assume $20\,\mathrm{mW}$. Based on the conditions in Eqs. (\ref{limits-1}) and (\ref{limits-2}), the control parameters, optimal potential $V_{ij}(t_k)$ and its time derivative acquire their maximum and minimum values which are summed up in some jump-wait sequences, where the jumps are characterized by the boundary values of $(\mathrm{d}/\mathrm{d}t)V_{ij}(t)$---determined according intensity modulator's bandwidth. We also note that restricting the laser pulse to a maximum value leads to some wait in the pulse shape. 

\begin{figure}[tp]
\centering
\includegraphics[width=1 \linewidth]{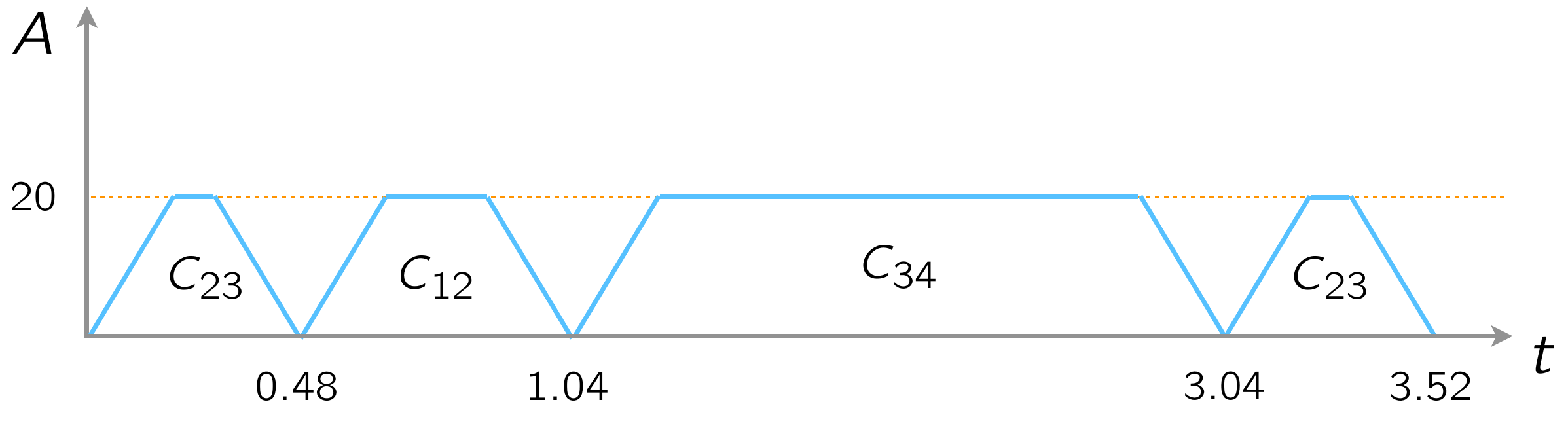}
\caption{Duration of pulses sequence ($\times 10^{-9}\,\mathrm{s}$) are calculated considering the maximum amplitude of each pulse ($\times \mathrm{mW}$) $20\,\mathrm{mW}$ and the modulation rate $0.1\,\mathrm{GHz}$.}
\label{fig-Pulses}
\end{figure}

Figure \ref{fig-Pulses} shows the result for pulse shapes and durations. 
By applying the pulse train, the system is totally in tunneling levels, thus tunneling is possible. Since tunneling is a probabilistic phenomena, the system should spend sufficient time in the unbounded tunneling levels in comparison to the transition time from the tunneling levels to the lower levels in this state to experience tunneling; otherwise, the system decays to the stable states and dissociation does not happen. Although the essential time for dissociation is unspecified in this method, we use less energy than the step size to decompose the composite system. However, this uncertainty in time is considerably small, as estimated above.
\section{Summary}
\label{sec:summary}

We have introduced a physically motivated and general definition for binding energy of bipartite quantum systems. In the making, we used a time-dependent potential to offset the interaction Hamiltonian and remove work-generating correlations between the subsystems. In this step, some physical considerations are taken into account. The potential $V(t)$ cannot be specified in general and is case dependent. For some systems, we may make the interaction Hamiltonian itself time-dependent to reset it to zero. If the system is endoergic, we need to spend a primary energy and, then, by passivating the system part of the spent energy returns to the agent. Finally, we have extended the definition of binding energy for probabilistic events, and through an example demonstrated that the probabilistic dissociation may be induced by quantum tunneling. 

\textit{Acknowledgements.}---This work was partially supported by Sharif University of Technology's Office of Vice President for Research and Technology through Contract No. QA960512 (to M.A. and F.B.).
F.B. also acknowledges support from the Ministry of Science, Research, and Technology of Iran (through funding for graduate research visits) and the Austrian Science Fund (FWF) through the START project Y879-N27 and the project P 31339- N27. 

%
\appendix

\section{Minimum dissociation time $t_{\mathrm{f}}$}
\label{App. Min time}

Our calculation is based on the group factorization and Pontryagin's maximum principle. The Lie group decomposition of a unitary operator can be employed to obtain the optimal control signal. There are several methods for group decomposition. Here we employ the planar rotation decomposition discussed in Ref. \cite{schirmer2002constructive}. A laser pulse of the following form is applied to the system:
\begin{align}
f_k(t)=2A_{k}(t)\cos(\omega_k t+\phi_k),
\end{align}
in which $A_k(t)$ is the pulse envelope, $\omega_k$ is the frequency of the transition $|\Phi_{k}^{[\mathrm{D}]}\rangle\rightarrow |\Phi_{k+1}^{[\mathrm{D}]}\rangle$. The system interacts with the applied laser field through its dipole moment, thus the interaction Hamiltonian (under some conditions) is given by
\begin{align}
H_k(t)= D_{kk} A_k(t)\big[e^{i(\omega_k t+\phi_k)}\ketbra{\Phi_{k}^{[\mathrm{D}]}}{\Phi_{k+1}^{[\mathrm{D}]}} + \mathrm{h.c.}],
\end{align}
where ``$\mathrm{h.c.}$" denotes Hermitian conjugate, and $D_{ij}= -e \langle \Phi_{i}^{[\mathrm{D}]}| \hat{x}| \Phi_{j}^{[\mathrm{D}]}\rangle$ is the dipole moment of the electron transition $|\Phi_{i}^{[\mathrm{D}]}\rangle \to |\Phi_{j}^{[\mathrm{D}]}\rangle$ caused by the laser pulse, with $e$ being the electron charge and $\hat{x}$ the position operator.

Let us consider the following anti-Hermitian matrices as a basis for the $su(d)$ Lie algebra:
\begin{align}
\hat{S}_{m,n}^{R}&=\ketbra{\Phi_{m}^{[\mathrm{D}]}}{\Phi_{n}^{[\mathrm{D}]}}-\ketbra{\Phi_{n}^{[\mathrm{D}]}}{\Phi_{m}^{[\mathrm{D}]}},\\
\hat{S}_{m,n}^{I}&=i(\ketbra{\Phi_{m}^{[\mathrm{D}]}}{\Phi_{n}^{[\mathrm{D}]}}+\ketbra{\Phi_{n}^{[\mathrm{D}]}}{\Phi_{m}^{[\mathrm{D}]}}),\\
\hat{S}_{m}&=\ketbra{\Phi_{m}^{[\mathrm{D}]}}{\Phi_{m}^{[\mathrm{D}]}}-\ketbra{\Phi_{m+1}^{[\mathrm{D}]}}{\Phi_{m+1}^{[\mathrm{D}]}},
\end{align}
where $1\leqslant  m \leqslant  d-1$ and $m \leqslant  n \leqslant  d$. It is straightforward to see that $\hat{X}_{k}:= \hat{S}_{k,k+1}^{R}$ and $\hat{Y}_{k}:=\hat{S}_{k,k+1}^{I}$, $1 \leqslant  k \leqslant  d$, suffice to generate the Lie algebra $\mathpzc{L}_0\subset su(d)$, which contains the generators $\hat{X}_{k}$ and $\hat{Y}_{k}$ for $1\leqslant  k \leqslant  d-1$. One can show that if the Lie algebra $\mathpzc{L}_0$ contains one of the pairs $(\hat{X}_1 , \hat{Y}_1)$ or $(\hat{X}_d ,\hat{Y}_d)$, then it must contains all the other generators. Using this, starting from any level in an atom, you can go up or down step-by-step to reach the desired level.

The sequences in which the fields should be turned on and off are obtained by decomposition of $\mathpzc{U}(t)$ into a product of generators of the dynamical Lie group, 
\begin{equation}
\label{decompose}
\mathpzc{U}(t) = \mathpzc{U}_{0}(t) \mathpzc{U}_{K} \mathpzc{U}_{K-1}\ldots \mathpzc{U}_{k}\ldots \mathpzc{U}_1.
\end{equation}
with $\mathpzc{U}_{0}(t)=e^{-itH}$. In the interaction picture and by applying the rotating-wave approximation, the interaction-picture Schr\"{o}dinger equation becomes
\begin{align*}
\frac{\partial \mathpzc{U}_{I}(t)}{\partial t} = \textstyle{\sum_{k=1}^{M}} A_{k}(t)\, D_{kk} [\hat{X}_{k} \sin{\phi_{k}} - \hat{Y}_{k} \cos{\phi_{k}}]\mathpzc{U}_I(t).
\end{align*}
Then if we apply in the interval $t_{k-1}\leqslant t \leqslant t_{k}$ a resonant pulse, then one can see that $\mathpzc{U}_{I}(t_{k})=\mathpzc{U}_{k}\mathpzc{U}_{I}(t_{k-1})$, where
\begin{align}
\label{decompose}
\mathpzc{U}_{k}= e^{C_{\sigma(k)} [\hat{X}_{\sigma(k)}\sin{\phi_{k}}- \hat{Y}_{\sigma(k)} \cos{\phi_{k}}]},
\end{align}
with
\begin{align}
\label{area}
C_{\sigma(k)}=D_{\sigma(k)\,\sigma(k)} \textstyle{\int_{t_{k-1}}^{t_{k}}} A_{\sigma(k)}(t)\,\mathrm{d}t
\end{align}
and $\sigma(k)$ being a mapping from the index set $\{1,\ldots,K\}$ to the control index set $\{1,\ldots,M\}$ that specifies the control $A_k$ which is on in the time interval $[t_{k-1},t_{k}]$. Here $K$ is the optimal number of the dipole transitions, and $M$ is the number of possible dipole transitions in the (composite) system.
 
 
\section{Optimal pulses for the example of Sec. \ref{subsec:tun}}
\label{App.example calculations}

As explained in the previous appendix, we need to decompose the unitary evolution operator into a product of unitary operators each of which is illustrative of a laser pulse which interacts with the dipole moment associated to a specific pair of consecutive levels of the Hamiltonian of the composite system. Each pulse is a $d\times d$ matrix (in the $\{|\Phi_{i}^{[\mathrm{D}]}\rangle\}_{i=1}^{d}$ basis)
with a nontrivial $2 \times 2$ block whose elements are specified by the dipole moments of the transitions $|\Phi_{k}^{[\mathrm{D}]}\rangle \to |\Phi_{k+1}^{[\mathrm{D}]}\rangle$, 
\begin{align}
\label{a-pulse}
\begin{pmatrix}
\mathbbmss{I}& & & & & & \\ \cline{3-4}
\multicolumn{2}{c|}{ } & \cos(C_{m}) & ie^{i\phi_{m}} \sin(C_{m}) & \multicolumn{2}{|c}{ } \\
\multicolumn{2}{c|}{ } & ie^{-i\phi_{m}}\sin(C_{m}) & \cos(C_{m}) & \multicolumn{2}{|c}{ } \\
 \cline{3-4} & & & & & & & \mathbbmss{I}
\end{pmatrix},
\end{align}
where $C_{m}$s are given by Eq. (\ref{area}) and $\phi_{m}$ is the phase of the pulse.

To find the pulse sequence, we shall follow the steps of the algorithm introduced in Ref. \cite{schirmer2002constructive}. Here the target unitary operator is given in Eq. (\ref{ev}), which can be written in the $\{|i\rangle\}_{i=1}^{4}$ basis as
\begin{align}
\label{unitarymatrix}
\mathpzc{U} =\begin{pmatrix}
0&0&1&0\\
0&0&0&1\\
1&0&0&0\\
0&1&0&0
\end{pmatrix}.
\end{align}
We now should find some unitary matrices of the form (\ref{a-pulse}) whose multiplication by $\mathpzc{U}$ results in the identity, $W_{K}\ldots W_{2}W_{1}\mathpzc{U}=\mathbbmss{I}$. In the first step, the last column of $\mathpzc{U}$ should be transformed to $(0\,0\,0\,1)^{T}$. This can be done by a pulse inducing the transition between levels $|2\rangle$ and $|3\rangle$, followed by another pulse between levels $|3\rangle$ and $|4\rangle$. The transition between levels $|2\rangle$ and $|3\rangle$ can be shown by a matrix of the form
\begin{align}
\label{a-pulse-}
W_{1}=\begin{pmatrix}
1 & & & \\
 &\cos(C_{1})&ie^{i\phi_{1}}\sin(C_{1})& \\
 &ie^{-i\phi_{1}}\sin(C_{1})&\cos(C_{1})& \\
 & & & 1
\end{pmatrix}.
\end{align}
To find the unknown parameters in these relations $C_{1}$ and $\phi_{1}$, we should use the column vector on which the pulse is applied. For example, consider the last column of Eq. (\ref{unitarymatrix}), for which we have
\begin{align}
W_{1}(a_1\,a_2\,a_3\,\,a_4)^{T} = (0\, 1\, 0\,0)^{T},
\label{aa-}
\end{align}
thus $a_2 =r_{2}e^{i\alpha_{2}}$ and $a_3=r_{3}e^{i \alpha_3}$ which leads to $C_{1}=-\cot^{-1}(-r_{3}/r_{2})$ and $\phi_{m}=\alpha_{2}-\alpha_{3}+\pi/2$.
Next, we should apply these two matrices on $\mathpzc{U}$ and look for some matrix that results in the following vector for the third column: $(0\,0\,1\,0)^{T}$. The rest of the pulse sequence may be calculated in a similar fashion, from which $\mathpzc{U}=W_{1}^{\dagger}\ldots W_{4}^{\dagger}$.

To find each pulse duration, due to Eq. \eqref{area}, the area covered by a pulse in an amplitude-time plot can be calculated,
\begin{align}
t_m-t'_m=C_m/(A_{m} D_{mm}),
\end{align}
in which $t_m$ is the pulse duration and $t'_m$ is the time for the laser pulse reaches its maximum---which is specified by the modulator. The optimal pulse shape has been represented in Fig. \ref{fig-Pulses}.

\end{document}